# On Applying Bandit Algorithm to Fault Localization Techniques

Masato Nakao
*Kindai University*
Higashi-osaka, Japan
2333340468p@kindai.ac.jp

Kensei Hamamoto
*Kindai University*
Higashi-osaka, Japan
2433340478s@kindai.ac.jp

Masateru Tsunoda
*Kindai University*
Higashi-osaka, Japan
tsunoda@info.kindai.ac.jp

Amjed Tahir
*Massey University*
Palmerston North, New Zealand
a.tahir@massey.ac.nz

Koji Toda
*Fukuoka Institute of Technology*
Fukuoka, Japan
toda@fit.ac.jp

Akito Monden
*Okayama University*
Okayama, Japan
monden@okayama-u.ac.jp

Keitaro Nakasai
*OMU College of Technology*
Osaka, Japan
nakasai@omu.ac.jp

Kenichi Matsumoto
*NAIST*
Ikoma, Japan
matumoto@is.naist.jp

*Abstract*—Developers must select a high-performance fault localization (FL) technique from available ones. A conventional approach is to try to select only one FL technique that is expected to attain high performance before debugging activity. In contrast, we propose a new approach that dynamically selects better FL techniques during debugging activity.

## I. Introduction

Fault localization (FL) [1] is an important process for identifying the course of faults or bugs in software. Assume that a developer identified a faulty module but was unable to locate the fault. FL helps identify the location of the faults in the source codes based on failed test cases, stack traces, or log reports. One key goal of FL is to help developers find faults faster and more efficiently. FL typically shows a list of candidates of faulty code snippets to developers.

There are various FL techniques. FL mainly consists of localization method and suspiciousness value. SBFL (Spectrum-based FL) and MBFL (mutation-based FL) are the two most known fault localization methods. For FL, the latter method uses artificial faults, and both methods use the suspiciousness value of each line of source code. Intuitively speaking, the suspiciousness value denotes the fault likelihood, and the suspiciousness value is based on the number of failed test cases. The suspiciousness value is calculated using various formulas (e.g., ochiai, tarantula, and dstar2 formulas) [1].

FL's performance (accuracy) can be different across the different techniques. Previous studies, such as [1], compared the performance of FL techniques to help developers select an appropriate technique and hence improve FL's performance. However, even if a particular technique attains high performance averagely on many software, the performance could degrade on certain software.

To help select FL techniques, we propose applying bandit algorithm (BA) [2]. BA is often explained through an analogy with slot machines. Assume that a player has 100 coins to bet on several slot machines. To maximize cumulated rewards, BA seeks sequentially best machines (referred to as arms) whose expected rewards are unknown to maximize cumulated rewards. We regard slot machines as FL techniques, and playing on a slot machine is like finding faults by a developer. That is, when the actual location is nearer to the fault location suggested by the selected arm, more coins (i.e., rewards) are acquired from the arm. The major differences from past studies such as [2] are the following:

- The application of BA to select FL techniques and evaluate the effect of BA.
- Use the average and median of a criterion as expected rewards of each arm and evaluate the influence on BA.

## II. Bandit Algorithm-Based Approach

**Procedure**: Based on the BA approach, developers perform the following steps to find and remove faults:

1. A developer picks up several FL techniques as arms.
2. One of the arms is selected based on the expected rewards of the arms.
3. The developer finds faults, referring to the location suggested by the selected arm.
4. The reward for each arm is calculated by comparing the actual location with the suggested one for each arm.
5. Expected rewards are updated based on the reward calculated in step 4.
6. Back to step 2.

Fig. 1 and 2 show the steps. Note that in step 2, one of the arms is randomly selected the first time. BA adds steps 1, 2, 4, and 5. Steps 2 and 5 can be automatically performed. Also, step 4 could be automated by inputting the actual locations of faults to an automation tool. As BA, we used ε-greedy, and set the parameter ε as 0. This is the simplest approach of BA, and it selects arms based on their expected rewards in step 2.

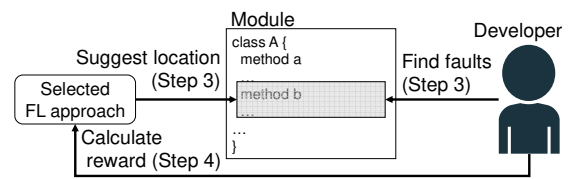

Fig. 1. **Procedure of FL based on BA**

Fig. 2. **Usage of EXAM score on BA**

TABLE I. EXAM SCORE OF EACH APROACH

| Evaluation Criterion | Measurement | Conventional (arm) | | | | BA | | Baseline |
|---|---|---|---|---|---|---|---|---|
| | | SBFL+ochiai | SBFL+tarantula | MBFL+ochiai | SBFL+dstar2 | BA-avg-EXAM | BA-mdn-EXAM | Random |
| Average EXAM | Rank | 7 | 2 | 1 | 4 | 3 | 5 | 6 |
| | score | 0.050049 | 0.010374 | 0.010026 | 0.011164 | 0.010653 | 0.013993 | 0.048509 |
| | Relative error | 80.0% | 3.4% | - | 10.2% | 5.9% | 28.4% | 79.3% |
| Median EXAM | Rank | 7 | 4 | 1 | 1 | 3 | 5 | 6 |
| | score | 0.009360 | 0.001072 | 0.000951 | 0.000951 | 0.000953 | 0.001078 | 0.006554 |
| | Relative error | 89.8% | 11.3% | - | - | 0.2% | 11.8% | 85.5% |

**Reward**: As a reward for step 2, we used the EXAM score, which is widely used to evaluate the performance of FL [1]. When the EXAM score is small, the performance is regarded as better. Intuitively speaking, the EXAM score denotes how much developers read source code to find defects by FL. The denominator of the score is the source lines of code of a target module, and the score is very different among modules. Therefore, as an evaluation criterion, median EXAM is used [1], in addition to the average, to avoid the influence of outliers.

If a developer focused on the median EXAM rather than the average, it is natural to use the median as the expected rewards in steps 2 and 5. Therefore, we used both average and median EXAM as the expected rewards. For instance, in Fig.2, when the average is used as the expected rewards of arms, MBFL is selected for the next target module. When the median is used, SBFL is selected. In what follows, we call the former approach *BA-avg-EXAM* and the latter one *BA-mdn-EXAM*.

Most of the past studies that applied BA to software fault prediction (but not to FL) [2] used AUC (area under the curve) as the expected reward of each arm. Very roughly speaking, AUC is a ratio, and therefore median and average cannot be calculated. That is, past studies did not clarify the influence of the difference in evaluation criteria.

## III. EXPERIMENT

**Dataset**: We used the dataset of Pearson et al., [1]. It includes results (i.e., EXAM scores) of various FL techniques. We selected cases where FL was applied to 133 modules of the Google Closure compiler. We randomly sorted the order of the results because BA could be affected by the order.

**Conventional approaches**: Study [1] evaluated various FL techniques. Based on the evaluation, as arms of BA, we picked up four techniques which were ranked as 0%, 25%, 75%, and 100 percentile by EXAM score (i.e., "MBFL + ochiai," "SBFL + tarantula," "SBFL + dstar2, " and "SBFL + ochiai.").

**Evaluation criteria**: We used the average and median EXAM score, the same as in the study [1]. That is, they are used as not only the expected reward of BA but also the evaluation criteria of FL, as shown in Fig 2. When evaluating BA, we applied it ten times. This is because BA selects an arm randomly on the first iteration (see Section 2).

Before applying FL, we could not decide on the best technique. Therefore, as the baseline, we randomly selected one of the techniques for 133 modules and calculated the average and median EXAM 10 times. To compare the techniques' performance easily, we defined relative error e = 1 – a / b. In the equation, b denotes the average or median EXAM of the best technique, and *a* does that of a target technique.

**Result of BA-avg-EXAM**: Table 1 shows the performance of FL techniques. The best one was MBFL + ochiai; therefore, we used its EXAM as the denominator of relative error. When we focus on average EXAM, SBFL + tarantula was the second best. In contrast, based on median EXAM, the rank of the technique was fourth, and the relative error of the median was larger than 10%. Similarly, SBFL + dstar2 was ranked first based on the median, but based on the average, the rank of the technique was lower, and the relative error of the average was larger than 10%.

Although BA-avg-EXAM was ranked third based on the average and the median, its relative error was smaller than 10%. Additionally, it was higher than SBFL + dstar2 based on the average and higher than SBFL + tarantula based on the median. Therefore, BA-avg-EXAM is effective for selecting FL techniques.

**Result of BA-mdn-EXAM**: In Table 1, although the performance of BA-mdn-EXAM was higher than the baseline, it was lower than BA-avg-EXAM. The result suggests that even when a developer focuses on median EXAM, avoiding using the median as an expected reward of arms might be good. In the table, the median EXAM of MBFL + Ochiai and SBFL + dstar2 was the same, which might make it challenging to select the best arms on early iterations of BA-mdn-EXAM.

**Discussion**: To find the best FL technique early, the number of evaluated FL techniques (i.e., rewards explained in Section 2) should be increased. One of the approaches is to use evaluations by other developers (e.g., other team members) who also apply BA to FL technique. The other approach is to utilize evaluations on similar modules or projects by contextual bandits, which accelerates the convergence, considering similarity.

## IV. CONCLUSION

We propose applying bandit algorithms (BA) to select a suitable fault localization technique. For the selection, we used the average and median of EXAM scores as expected rewards of BA. Our experiment showed that BA's accuracy was higher than a baseline, and using the average EXAM was more accurate than the median EXAM.

REFERENCES

[1] S. Pearson et al., "Evaluating and Improving Fault Localization," Proc. International Conference on Software Engineering (ICSE), pp.609-620, 2017.

[2] M. Tsunoda et al., "Using Bandit Algorithms for Selecting Feature Reduction Techniques in Software Defect Prediction," Proc. International Conference on Mining Software Repositories (MSR), pp.670-681, 2022.